\documentclass[12pt,a4paper]{article}
\usepackage{amsmath,amssymb}
\usepackage{epsfig}

\begin{document}
\begin{center}
\Large\textbf{\uppercase{ Problem of cosmological singularity and
inflationary cosmology}}\\
\normalsize A. V. Minkevich\\
 \textit{$^1$Department of Theoretical Physics, Belarussian State University, Minsk, Belarus\\
 $^2$Department of Physics and Computer Methods, Warmia and Mazury University in Olsztyn,
 Poland\\
\normalsize  E-mail: MinkAV@bsu.by; awm@matman.uwm.edu.pl}
\end{center}
\begin{flushright}
\begin{minipage}{0.8\textwidth}
\textbf{Abstract.} Problem of cosmological singularity of general
relativity theory is discussed. The possible resolution of this
problem in the framework of inflationary cosmology is proposed.
Physical conditions leading to bouncing inflationary solutions in
the frame of general relativity theory and gauge theories of
gravitation are compared. It is shown that gauge theories of
gravitation allow to build regular inflationary cosmological
models of closed, open and flat type with dominating
ultrarelativistic matter at a bounce.
\end{minipage}
\end{flushright}

\section{Problem of cosmological singularity and vacuum\\
gravitational repulsion effect}

Problem of cosmological singularity (PCS) is one of the most
principal problems of general relativity theory (GR). According to
Hawking-Penrose theorems the appearance of cosmological
singularity in cosmological solutions of GR is inevitable, if
gravitating matter satisfies so-called energy dominance conditions
[1, 2]. There were many attempts to resolve PCS. If the Planckian
(pre-Planckian) epoch took place in the beginning of cosmological
expansion, consequent quantum theory of gravitation is necessary
to analyze the PCS\footnote{Recently the PCS was discussed in the
frame of string theory (see [3, 4] and refs given here).}. But
such theory does not exist at present. By supposing that the
Planckian epoch was absent by evolution of the Universe, classical
theory of gravitation (with quantum description of gravitating
matter)can be used by analysis of the PCS. There were many such
attempts to resolve the PCS corresponding to the following two
approches (see [5] and refs given here).

In the frame of the first approach one supposes that energy
dominance conditions are valid for gravitating matter with
extremely high energy densities and pressures at the beginning of
cosmological expansion. In this case GR must be replaced by
certain other gravitational theory by description of gravitating
matter at such extreme states. Gauge theories of gravitation (GTG)
and at first of all the Poincare GTG are an important
generalization of GR. Cosmological equations for homogeneous
isotropic models deduced in the frame of GTG lead to regular in
metrics cosmological solutions in the case of usual gravitating
matter [6]. These solutions differ from corresponding solutions of
GR only at extreme states of gravitating matter near certain
limiting energy density, where gravitational interaction has the
character of repulsion instead of attraction.

The second approach to resolve PCS concerns the analysis of
description of gravitating matter at extreme conditions and
examination of energy dominance conditions for corresponding
states\footnote{Such analysis is important also in connection with
the problem of dark matter and dark energy [7].}. Unified gauge
theories of strong and electroweak interactions with spontaneous
symmetry breaking are the modern base of matter description at
extreme conditions.  Inflationary cosmology as important part of
the theory of early Universe was built by using gauge theories of
elementary particles [8,9]. A number of problems of standard
Friedmann cosmology were resolved in the frame of inflationary
cosmology. Most inflationary cosmological models discussed in
literature are singular and their study is given from Planckian
time. The radical idea of quantum birth of the Universe was
introduced in order to avoid the PSC. Note that during
quasi-de-Sitter inflationary stage gravitating matter in
inflationary models is approximately in the state of so-called
gravitating vacuum, for which pressure $p$ and energy density
$\rho>0$ are connected in the following way $p=-\rho$ and energy
dominance conditions are not valid. As it was shown in Refs. [10,
11] gravitating vacuum with sufficiently large energy density can
lead to the vacuum gravitational repulsion effect (VGRE) in the
case of systems including also usual gravitating matter, that
allows to build regular inflationary cosmological models. At first
the VGRE was discussed in the frame of Poincare GTG [10] in the
case of homogeneous isotropic models including radiation and
gravitating vacuum with $\rho=\mathrm{const}$. Because
cosmological equations of Poincare GTG are valid in the frame of
the most general GTG --- metric-affine GTG, the VGRE can take
place also in metric-affine GTG [12, 13]. Regularizing role of
gravitating vacuum in GTG was analyzed in Refs [14-17]. The VGRE
in GTG can lead to a bouncing solutions in the case of closed,
flat and open models. It is important that the VGRE can lead to
the bounce also in GR in the case of closed models [11, 5] (see
also [26] and refs given here). Regular bouncing closed models
filled by linear massive scalar field were studied in the frame of
GR in Refs. [18, 19]. By using values of some parameters of the
Universe, the probability of regular solutions for such models was
estimated to be very small [18]\footnote{At the first time
cosmological solution with initial de Sitter state limited in the
past was discussed in Ref. [20]. The problems - what origin has
initial vacuum state, what was before this state - were not
considered in [20]. The cosmological solution with initial vacuum
state not limited in the time in the past, proposed in Ref.[21]
does not lead to resolution of PCS in the future.}. This
estimation is not valid for inflationary models discussed below,
for which scalar fields dominate during very small time intervals.

By using some simplest potentials for scalar fields regular
inflationary cosmological models were discussed in the frame of GR
as well as GTG in Refs.[22-25]. It was shown that limiting energy
density and temperature at the bounce in inflationary models can
be essentially smaller than the Planckian ones. Obtained regular
inflationary solutions contain the stage of transition from
compression to expansion and quasi-de-Sitter inflationary stage.
After inflation scalar fields oscillate and are transformed into
elementary particles, as result the transition to Friedmann regime
takes place. Formal integration of cosmological equations leads
also to quasi-de-Sitter stage of compression [22-25]. However,
this stage in GR is unstable and small change of variables at
compression stage can lead to singular solution. This means that
the building of regular inflationary models is possible, if
physical conditions leading to the bounce take place at the end of
cosmological compression (see below).

By using results of numerical investigation performed in Ref.[27],
the analysis of regular inflationary models in GR and GTG in
dependence on conditions at a bounce is given below in present
paper.

\section{Regular inflationary cosmological models in GR}

Let us consider homogeneous isotropic models filled by scalar
field $\phi$ minimally coupled with gravitation and usual matter.
We suppose that interaction between them is negligibly small. Then
energy density $\rho$ and pressure $p$ can be written in the
following form
\begin{equation}
\rho=\frac{1}{2}\dot{\phi}^2+V(\phi)+\rho_m, \qquad
p=\frac{1}{2}\dot{\phi}^2-V(\phi)+p_m,
\end{equation}
where a dot denotes differentiation with respect to time, index
$m$ corresponds to usual matter, and $V(\phi)$ is scalar field
potential. Evolution of considered models in GR is described by
Friedmann cosmological equations
\begin{equation}
\frac{k}{R^2}+H^2 =\frac{8
\pi}{3M_p^2}\,\left(\frac{1}{2}\dot{\phi}^2+V(\phi)+\rho_m\right),
\end{equation}
\begin{equation}
\dot{H}+H^2 =\frac{8
\pi}{3M_p^2}\,\left(V(\phi)-\dot{\phi}^2-\frac{1}{2}\left(\rho_m+3p_m\right)\right),
\end{equation}
where $k=+1,0,-1$ for closed, flat  and open models respectively,
$R$ is the scale factor,  $H=\dot{R}/R$ is Hubble parameter, $M_p$
is Planckian mass. (The system of units with $\hbar=c=1$ is used.)
By virtue of negligible interaction of scalar field with usual
matter the conservation law
\begin{equation}
\dot{\rho}+3H\left(\rho+p\right)=0
\end{equation}
leads to equation for scalar field
\begin{equation}
\ddot{\phi}+3H\dot{\phi}=-V' \qquad
\left(V'=\frac{dV}{d\phi}\right)
\end{equation}
and
\begin{equation}
\dot{\rho}_m+3H(\rho_m+p_m)=0.
\end{equation}
In the case of closed models (k=+1) Eqs. (2)--(3) can lead to
regular bouncing solutions. Putting $t=0$ at bounce ($H(0)=0$,
$\dot{H}_0\equiv\dot{H}(0)>0$), we see bouncing solutions take
place, if the value of scalar field potential $V_0=V(\phi(0))$
(which plays the role of gravitating vacuum energy density) is
sufficiently large and satisfies the following inequality
\begin{equation}
V_0>\dot{\phi}_0^2+\frac{1}{2}\left(\rho_{m0}+3p_{m0}\right)
\end{equation}
(index $0$ corresponds to $t=0$). For definiteness we will
consider later the matter in the form of ultrarelativistic matter
or radiation ($\rho_m=\rho_r$, $p_m=\frac{1}{3}\rho_r$, index $r$
corresponds to radiation), then Eq. (3) takes the following form
$$
 \dot{H}+H^2 = \frac{8 \pi}{3M_p^2}\,
\left(V-\dot{\phi}^2-\rho_{r} \right).\eqno(3')
$$
and according to (6) we have
\begin{equation}
\rho_rR^4={\rm const}.
\end{equation}

To obtain a bouncing solution we have to take some initial
conditions for scalar field ($\phi_0$, $\dot{\phi}_0$) and
radiation $\rho_{r0}$ satisfying according to (7) the following
relation
$$
V_0-\dot{\phi}_0^2-\rho_{r0}>0\eqno(7')
$$
Admissible values of $\phi_0$, $\dot{\phi}_0$ and $\rho_{r0}$ must
satisfy restrictions: $V_0\lesssim 1 M_p^4$,
$\frac{1}{2}\dot{\phi}_0^2 \lesssim 1M_p^4$, $\rho_{r0}\lesssim 1
M_p^4$, by which quantum gravitational effects are not essential
and our classical consideration is valid. Minimum value of $R_0$
of scale factor at bounce we obtain from Eq. (2)
\begin{equation}
R_0=\left[\frac{8 \pi}{3M_p^2}\,
\left(V_0+\frac{1}{2}\dot{\phi}_0^2+\rho_{r0}
\right)\right]^{-\frac{1}{2}}
\end{equation}
By using initial conditions at bounce we integrate Eqs. (3$'$) and
(5) putting $\displaystyle \rho_r=\frac{\rho_{r0} R_0^4}{R^4}$ in
accordance with (8).

Now let us discuss the most important features of regular
inflationary models in GR, by taking into account numerical
results obtained in the case of the simplest scalar field
potentials $V_1=\frac{1}{4}\lambda\phi^4$ ($\lambda=10^{-14}$) and
$V_2=\frac{1}{2}m^2\phi^2$ in Ref. [25].

\medskip
a) Transition stage from compression to expansion\\

At first we will consider bouncing solutions in the case when the
value of $\eta\equiv V_0-\dot{\phi}_0^2-\rho_{r0}>0$ is not near
to zero. Then as numerical analysis shows, the Hubble parameter
$H(t)$ and scalar field $\phi(t)$ ($\dot{\phi}_0\ne 0$) vary in
linear way during transition stage from compression to expansion,
and derivatives $\dot{H}$ and $\dot{\phi}$ decrease rapidly and
tend to zero at the end of this stage ($t\sim t_1$), moreover
radiation energy density $\rho_r\to 0$ at $t\sim t_1$ (see Fig.
1).\footnote{All figures in this paper are given in system of
units $\hbar=c=M_p=1$ for the potential $V=\frac{1}{4}\lambda
\phi^4$ ($\lambda=10^{-14}$).}
\begin{figure}[htb!]
\begin{minipage}{0.48\textwidth}\centering{
\epsfig{file=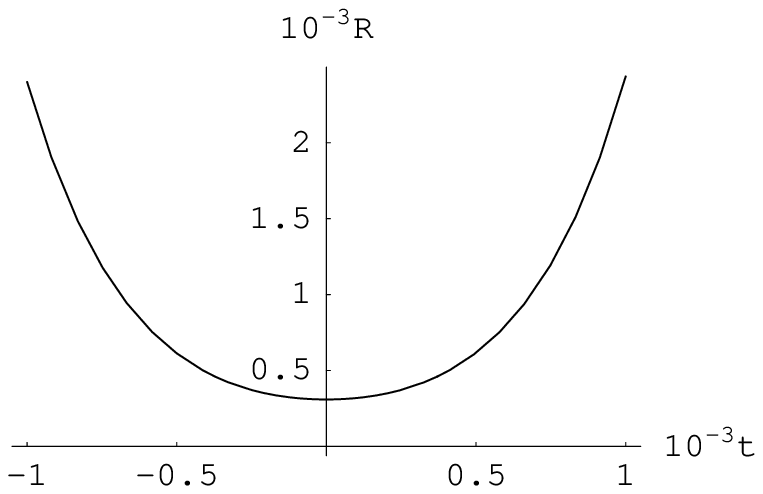,width=\linewidth}}
\end{minipage}\, \hfill\,
\begin{minipage}{0.48\textwidth}\centering{
\epsfig{file=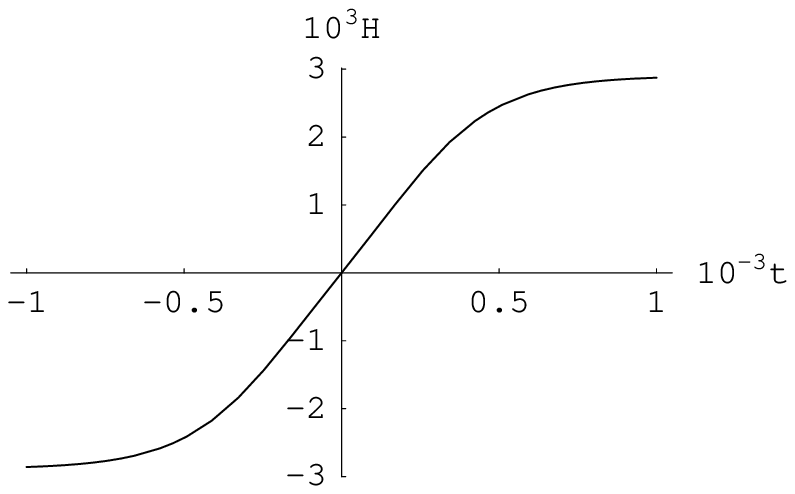,width=\linewidth}}
\end{minipage}\\
\begin{minipage}{0.48\textwidth}\centering{
\epsfig{file=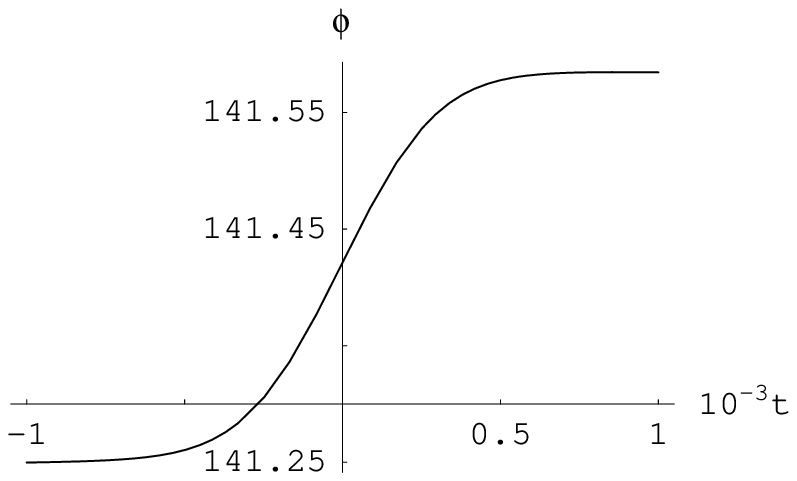,width=\linewidth}}
\end{minipage}\, \hfill\,
\begin{minipage}{0.48\textwidth}\centering{
\epsfig{file=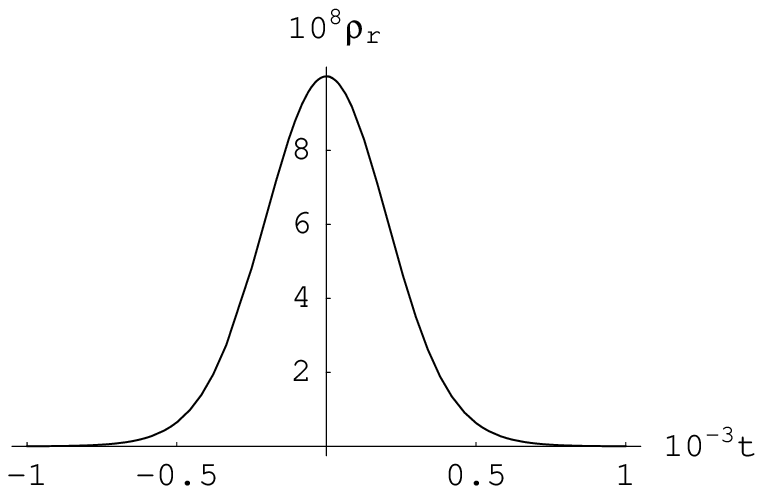,width=\linewidth}}
\end{minipage}
\caption[]{Dynamics of functions $R(t)$, $H(t)$, $\phi(t)$ and
$\rho(t)$ during transition stage from compression to expansion in
the case of  initial conditions: $V_0=10^{-6}M_p^4$
($\phi_0=141.42M_p$), $\dot{\phi}_0=\sqrt{0.3V_0}$,
$\rho_{r0}=0.1V_0$}
\end{figure}
According to Eq. (3$'$) we have
\begin{equation}
H(t_1)=\sqrt{\frac{8 \pi}{3M_p^2}\,
V(\phi_1)}\qquad(\phi_1=\phi(t_1))
\end{equation}
and the change of Hubble parameter during time interval $\Delta
t_1=t_1-t_0$ ($t_0=0$, $H_0=0$) is
\begin{equation}
\Delta H_1=H_1=\dot{H}_0 \Delta t_1=\frac{8
\pi}{3M_p^2}\,\left(V_0-\dot{\phi}_0^2-\rho_{r0}\right)\, \Delta
t_1.
\end{equation}
Because $\phi_1=\phi_0+\dot{\phi}_0 \Delta t_1$ and hence
according to (10) $\displaystyle H_1=\sqrt{\frac{8
\pi}{3M_p^2}V_0}\left(1+\frac{1}{2}\frac{V_0'\dot{\phi}_0}{V_0}\Delta
t_1\right)$ we find from (11)
\begin{equation}
\Delta t_1=\frac{\displaystyle\sqrt{V_0}}{\displaystyle
\sqrt{\frac{8 \pi}{3M_p^2}}\,
\left(V_0-\dot{\phi}_0^2-\rho_{r0}\right)-\frac{1}{2}\frac{V_0'}{\sqrt{V_0}}
\dot{\phi}_0}
\end{equation}
Analogously we can determine time interval $\Delta t_2$ of
transition stage before a bounce
$$
\Delta t_2=\frac{\displaystyle\sqrt{V_0}}{\displaystyle
\sqrt{\frac{8 \pi}{3M_p^2}\,}\;
\left(V_0-\dot{\phi}_0^2-\rho_{r0}\right)+\frac{1}{2}\frac{V_0'}{\sqrt{V_0}}
\dot{\phi}_0}
$$
Then duration of transition stage is
\begin{equation}
\Delta t_{tr}=\Delta t_1+\Delta t_2=\frac{\displaystyle
2\;\sqrt{\frac{8 \pi}{3M_p^2}\,V_0\,}\:
\left(V_0-\dot{\phi}_0^2-\rho_{r0}\right)}{\displaystyle \frac{8
\pi}{3M_p^2}\,
\left(V_0-\dot{\phi}_0^2-\rho_{r0}\right)^2-\frac{1}{4}\frac{V_0'^2}{V_0}
\dot{\phi}_0^2}
\end{equation}
and value of the Hubble parameter at the end of transition stage
(at the beginning of inflationary stage) is
\begin{equation}
H_1=\frac{\displaystyle \frac{8
\pi}{3M_p^2}\,\left(V_0-\dot{\phi}_0^2-\rho_{r0}\right)
\sqrt{V_0}}{\displaystyle \sqrt{\frac{8 \pi}{3M_p^2}}\,
\left(V_0-\dot{\phi}_0^2-\rho_{r0}\right)-\frac{1}{2}\frac{V_0'}{\sqrt{V_0}}
\dot{\phi}_0}.
\end{equation}
If $\displaystyle \frac{1}{2}\left|\frac{V_0'}{\sqrt{V_0}}
\dot{\phi}_0\right|\ll \sqrt{\frac{8 \pi}{3M_p^2}}\,
\left(V_0-\dot{\phi}_0^2-\rho_{r0}\right)$, we have
\begin{equation}
H_1\approx\sqrt{\frac{8 \pi}{3M_p^2}\,V_0}.
\end{equation}
The value of $H_1$ (15) depends on $\phi_0$ only and does not
depend on $\dot{\phi}_0$ and $\rho_{r0}$.

By decreasing of value of $\eta$, the linear time dependence of
$H(t)$ is violated and duration of transition stage increases (see
Fig. 2). Formulas (13)--(14) are not applicable in this case, but
expression (10) for the Hubble parameter at the end of transition
stage is valid.
\begin{figure}[htb!]
\hfill\, \begin{minipage}{0.6\textwidth}\centering{
\epsfig{file=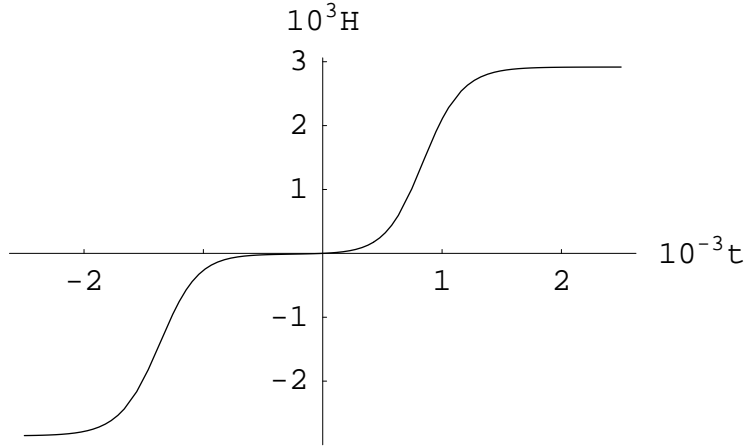,width=\linewidth}} \caption[]{Dynamics of
function  $H(t)$ during transition stage from compression to
expansion in the case of initial conditions: $V_0=10^{-6}M_p^4$,
$\dot{\phi}_0=\sqrt{0.5V_0}$, $\rho_{r0}=0.49V_0$}
\end{minipage}\hfill ,
\end{figure}

In the case of very small values of $\eta\to 0$ solutions change
their character --- inflationary stage and oscillations of scalar
field vanish, solutions become unstable. For example, in the case
of potential $V=\frac{1}{4}\lambda \phi^4$ ($\lambda=10^{-14}$) we
have inflationary solutions for initial conditions: a)
$V_0=10^{-14}M_p^4$, $\dot{\phi}_0=0$, $\rho_{r0}=0$; b)
$V_0=0.1M_p^4$, $\dot{\phi}_0=0$, $\rho_{r0}=(1-10^{-6})V_0$; c)
$V_0=0.1M_p^4$, $\dot{\phi}_0=\sqrt{(1-10^{-3})V_0}$,
$\rho_{r0}=0$. However, we don't have inflationary solutions by
the following change of these conditions: a) $V_0=10^{-15}M_p^4$
($\phi_0=0.8 M_p$), $\dot{\phi}_0=0$, $\rho_{r0}=0$; b)
$V_0=0.1M_p^4$, $\dot{\phi}_0=0$, $\rho_{r0}=(1-10^{-7})V_0$; c)
$V_0=0.1M_p^4$, $\dot{\phi}_0=\sqrt{(1-10^{-4})V_0}$,
$\rho_{r0}=0$.

b) Inflationary stage

During quasi-de-Sitter inflationary stage we have $\dot{H}\ll
H^2$, $\dot{\phi}^2\ll V(\phi)$, $\ddot{\phi}\ll V'$,
$\rho_r\approx 0$ and term $R^{-2}$ in Eq. (2) is negligibly
small. Then according to Eq. (3$'$) (and Eq. (2)) the Hubble
parameter at inflationary stage is equal to
\begin{equation}
H=\frac{d\log R}{dt}=\sqrt{\frac{8 \pi}{3M_p^2}V(\phi)}
\end{equation}
and Eq. (5) for scalar field takes form
\begin{equation}
\dot{\phi}=-\sqrt{\frac{M_p^2}{6\pi}}\, \frac{d\sqrt{V}}{d\phi}.
\end{equation}
Values of scalar field $\phi_1=\phi(t_1)$ and scale factor
$R_1=R(t_1)$ play the role of initial conditions by integration of
Eqs. (16)--(17) for inflationary stage. For given potential
$V(\phi)$ the integration of Eq. (17) leads to the function
$\phi(t)$. In the case of potential $V_1=\frac{1}{4}\lambda
\phi^4$ we obtain
\begin{equation}
\phi(t)=\phi_1 e^{\textstyle -\sqrt{\frac{\lambda}{6 \pi}} M_p
\left(t-t_1\right)}
\end{equation}
and in the case of potential $V_2=\frac{1}{2} m^2 \phi^2$ we have
\begin{equation}
\phi(t)=\phi_1-\frac{M_p}{2\sqrt{3\pi}}\, m\left(t-t_1\right).
\end{equation}
By using explicit form of scalar field $\phi$ we can find the
scale factor by integrating Eq. (16). In the case of functions
(18)--(19) we obtain
\begin{equation}
R(t)=R_1\exp\left[\frac{4\pi}{n
M_p^2}\left(\phi_1^2-\phi^2(t)\right)\right],
\end{equation}
where $n=4$ and $n=2$ in the case of potentials $V_1$ and $V_2$
respectively. Expressions (18)--(20) have the same form as in
chaotic inflation [8], but in considered theory the values
$\phi_1$ and $R_1$ are determined by initial conditions at bounce
and are not arbitrary.

At the end of inflationary stage Eqs. (16)--(17) are not valid,
and behaviour of scalar field and Hubble parameter can be
determined from Eqs. (3$'$) and (5). After inflation the Hubble
parameter is small and decreases, and scalar field changes as
damped oscillations (Fig. 3).
\begin{figure}[htb!]
\begin{minipage}{0.48\textwidth}\centering{
\epsfig{file=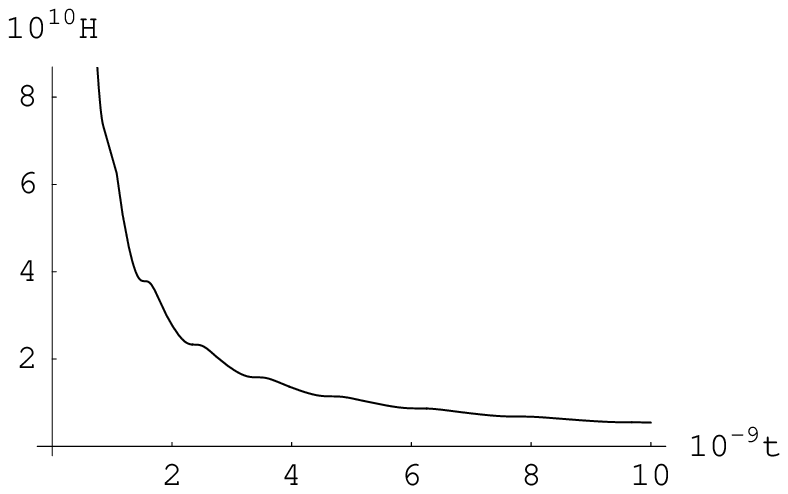,width=\linewidth}}
\end{minipage}\, \hfill\,
\begin{minipage}{0.48\textwidth}\centering{
\epsfig{file=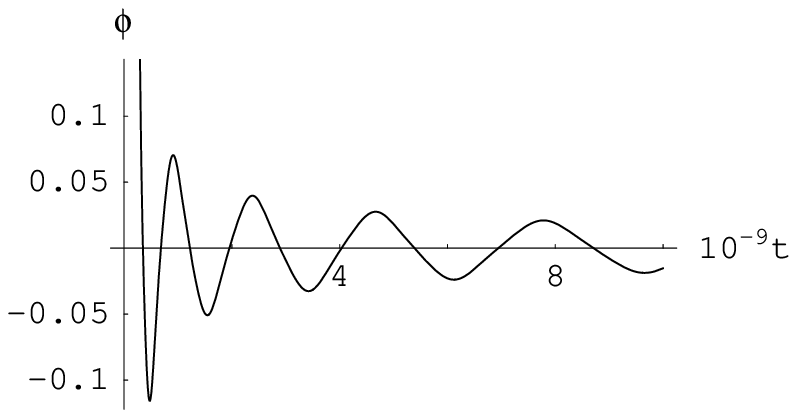,width=\linewidth}}
\end{minipage}
\caption[]{Dynamics of functions  $H(t)$ and $\phi(t)$ after
inflation ($V_0=10^{-6}M_p^4$, $\dot{\phi}_0=\sqrt{0.3V_0}$,
$\rho_{r0}=0.1V_0$).}
\end{figure}
Oscillations characteristics (amplitude, frequency) depend on
potential parameters and practically do not depend on initial
conditions at bounce. Such dependence appears in the case of
bouncing solutions with small values of $\eta$ near their
stability boundary [27]. But noted solutions do not describe
inflationary models and are not interesting for inflationary
cosmology. Note, that relative part of such solutions in
comparison with bouncing inflationary solutions is small. As
result approximately relative part of bouncing inflationary
solutions on the plane ($\phi_0$,  $\dot{\phi}_0$) satisfying
inequality (7$'$) (if $\rho_{r0}=0$, $0\le V(\phi_0)\le 1 M_p^4$,
$0\le \frac{1}{2}\dot{\phi}_0^2\le 1 M_p^4$) is equal to:
a)$\frac{\sqrt{2}}{6}$ for potential
$V_1=\frac{1}{4}\lambda\phi^4$, b) $\frac{\sqrt{2}}{4}$ for
potential $V_2=\frac{1}{2}m^2\phi^2$ [25].

\section{Regular inflationary cosmological models in GTG}
Cosmological equations for homogeneous isotropic models obtained
in the frame of gauge theories of gravitation have the following
form
\begin{equation}
\frac{k}{R^2}+\left\{\frac{d}{dt}\ln\left[R\sqrt{\left|1-\beta\left(\rho-
3p\right)\right|}\,\right]\right\}^2=\frac{8\pi}{3M_p^2}\,\frac{\rho-
\frac{\beta}{4}\left(\rho-3p\right)^2}{1-\beta\left(\rho-3p\right)}
\, ,
\end{equation}
\begin{equation}
\frac{\left[\dot{R}+R\left(\ln\sqrt{\left|1-\beta\left(\rho-
3p\right)\right|}\,\right)^{\cdot}\right]^\cdot}{R}=
-\frac{4\pi}{3M_p^2}\,\frac{\rho+3p+\frac{\beta}{2}\left(\rho-3p\right)^2}{
1-\beta\left(\rho-3p\right)}\, .
\end{equation}
Here $\beta$ is indefinite parameter with inverse dimension of
energy density. At first Eqs. (21)--(22) were deduced in Poincare
GTG [6], and later it was shown that Eqs. (21)--(22) take place
also in metric-affine GTG [12, 13]. The conservation law has usual
form (4). Besides cosmological equations (21)--(22) gravitational
equations of GTG lead to the following relation for torsion
function $S$ and nonmetricity function $Q$
\begin{equation}
S-\frac{1}{4}Q=-\frac{1}{4}\,\frac{d}{dt}
\ln\left|1-\beta(\rho-3p)\right|.
\end{equation}
In Poincare GTG $Q=0$ and Eq. (23) determines the torsion
function. In metric-affine GTG there are three kinds of models
[13]: in the Riemann-Cartan space-time ($Q=0$), in the Weyl
space-time ($S=0$), in the Weyl-Cartan space-time ($S\neq 0$,
$Q\neq 0$, the function S is proportional to the function $Q$).
Eqs. (21)--(22) coincide practically with Friedmann cosmological
equations of GR and noneinsteinian characteristics $S$ and $Q$ are
negligible if energy density is small
$\left|\beta(\rho-3p)\right|\ll 1$. The difference between GR and
GTG can be essential at extremely high energy densities
$\left|\beta(\rho-3p)\right|\gtrsim 1$.\footnote{Ultrarelativistic
matter with equation of state $p=\frac{1}{3}\rho$ is exceptional
system because Eqs. (21)--(22) are identical to Friedmann
cosmological equations of GR in this case.} In the case of
gravitating vacuum with constant energy density
$\rho_v=\mathrm{const}>0$ cosmological Eqs. (21)--(22) are reduced
to Friedmann cosmological equations of GR and $S=Q=0$, this means
that de Sitter solutions for metrics with vanishing torsion and
nonmetricity are exact solutions of GTG [28,29] and hence
inflationary models can be built in the frame of GTG [22-24]. In
the case of values of $\beta$ to be not large in system of units
with $\hbar=c=M_p=1$ ($\left|\beta(4V-\dot{\phi}^2)\right|\ll 1$)
inflationary models of GTG practically coincide with that of GR
[24]. It is interesting to investigate inflationary models in GTG
in the case of large values of $\beta$
($\left|\beta(4V-\dot{\phi}^2)\right|\gg 1$).

In order to analyze inflationary cosmological models in GTG let us
consider systems including scalar field and radiation, for which
quantities $\rho$ and $p$ are given by (1), where $\rho_m=\rho_r$,
$p_m=\frac{1}{3}\rho_r$. Then cosmological equations (21)--(22) by
taking into account Eq. (5) can be written in the following form
\begin{eqnarray}
& & \frac{k}{R^2}
Z^2+\left\{H\left[1-2\beta(2V+\dot{\phi}^2)\right]-3\beta
V'\dot{\phi}\right\}^2\nonumber \\
& & \phantom{+H\left[1-2\beta(2V+\dot{\phi}^2)\right]} =\frac{8
\pi}{3M_p^2}\,\left[\rho_r+
\frac{1}{2}\dot{\phi}^2+V-\frac{1}{4}\beta\left(4V-\dot{\phi}^2
\right)^2\right]Z,\\
& &\dot{H}\left[1-2\beta(2V+\dot{\phi}^2)\right]Z+H^2\left\{\left[
1-4\beta(V-4\dot{\phi}^2)\right]Z-18\beta^2\dot{\phi}^4\right\}\nonumber\\
& &\phantom{H} +12\beta
H\dot{\phi}V'\left[1-2\beta(2V+\dot{\phi}^2)\right]-
3\beta\left[(V''\dot{\phi}^2-V'{}^2)Z+6\beta\dot{\phi}^2
V'{}^2\right]
\nonumber\\
& &\phantom{\dot{H}\left[1-2\beta(2V+\dot{\phi}^2)\right]Z}
=\frac{8 \pi}{3M_p^2}\,\left[V-\dot{\phi}^2-\rho_r
-\frac{1}{4}\beta(4V-\dot{\phi}^2)^2\right]Z,
\end{eqnarray}
where $Z=1-\beta(4V-\dot{\phi}^2)$, $\displaystyle
V'=\frac{dV}{d\phi}$, $\displaystyle V''=\frac{d^2V}{d\phi^2}$.
Relation (23) takes the form
\begin{equation}
S-\frac{1}{4}Q=\frac{3\beta}{2}\,\frac{\left(H\dot{\phi}+V'\right)\dot{\phi}}{1-\beta
\left(4V-\dot{\phi}^2\right)}.
\end{equation}
Eqs.(24)-(25) lead to essential restrictions on scalar field by
evolution. So in the case $k=0,+1$ from Eq.(24) follows that
$Z\geq 0$, if $\beta<0$. Because at the bounce ($t=0$) we have
$H(0)=0$, from (24) follows
\begin{equation}
\frac{k}{R_0^2} Z_0^2+9\beta^2 V'{}_0^2\dot{\phi}_0^2=\frac{8
\pi}{3M_p^2}\,\left[\rho_{r0}+\frac{1}{2}\dot{\phi_0}^2
+V_0-\frac{1}{4}\beta\left(4V_0-\dot{\phi}_0^2\right)^2\right]Z_0
\end{equation}
($Z_0=1-\beta(4V_0-\dot{\phi}_0^2)$). The relation (27) determines
minimum value of scale factor $R_0$ for closed and open models,
and in the case of flat models gives the dependence between
$\phi_0$, $\dot{\phi}_0$ and $\rho_{r0}$. From (25) the time
derivative $\dot{H}_0=\dot{H}(0)$ at the bounce is
\begin{eqnarray}
& &\dot{H}_0=\left\{\frac{8
\pi}{3M_p^2}\,\left[V_0-\dot{\phi}_0^2-\rho_{r0}-\frac{1}{4}\beta(4V_0
-\dot{\phi}_0^2)^2\right]Z_0\right.\nonumber\\
&
&\left.\phantom{H}+3\beta\left[(V''{}_0\dot{\phi}_0^2-V'{}_0^2)Z_0
+6\beta\dot{\phi}_0^2 V'{}_0^2\right]\right\}
\left[1-2\beta(2V_0+\dot{\phi}_0^2)\right]^{-1}Z_0^{-1}.
\end{eqnarray}
Then the condition $\dot{H}_0>0$ determines permissible values of
$\phi_0, \dot{\phi}_0, \rho_{r0}$ at the bounce in dependence on
indefinite parameter $\beta$. In order to analyze the dependence
$\dot{H}_0=\dot{H}_0(\beta)$ we write expression (28) in the form
\begin{equation}\label{eq29}
\dot{H}_0=a\beta_1\beta_2\left(\beta^2+\frac{b}{a}\beta+\frac{f}{a}\right)
\left(\beta-\beta_1\right)^{-1}\left(\beta-\beta_2\right)^{-1},
\end{equation}
where $\beta_1=\frac{1}{2}(2V_0+\dot{\phi}_0^2)^{-1}$ and
$\beta_2=(4V_0-\dot{\phi}_0^2)^{-1}$ are particular points and
\begin{align*}
a=&\frac{2\pi}{3M_p^2}\,\left(4V_0-\dot{\phi}_0^2\right)^{3}+
3\left(V_0'^2-V_0''\dot{\phi}_0^2\right)\left(4V_0-\dot{\phi}_0^2\right)
+18V_0'^2\dot{\phi}_0^2,\\
b=&\frac{2\pi}{3M_p^2}\,\left(4V_0-\dot{\phi}_0^2\right)\,
\left(4\rho_{r0}+5\dot{\phi}_0^2-8V_0\right)-3\left(V_0'^2-V_0''\dot{\phi}_0^2\right),\\
f=&\frac{8\pi}{3M_p^2}\,\left(V_0-\dot{\phi}_0^2-\rho_{r0}\right).\\
\end{align*}
If $b^2-4a f\ge0$, the formula (29) takes the form
$$
\dot{H}_0=a\beta_1\beta_2
\left(\beta-\beta_1^{(0)}\right)\left(\beta-\beta_2^{(0)}\right)
\left(\beta-\beta_1\right)^{-1}\left(\beta-\beta_2\right)^{-1},
$$
where $\displaystyle \beta_{1,2}^{(0)}=\frac{-b\pm \sqrt{b^2-4a
f}}{2a}$. In the case of large (in module) values of $\beta$
($|\beta\left(4V_0-\dot{\phi}_0^2\right)|\gg 1$) we have
$\dot{H}_0\approx a\beta_1\beta_2$ and the bounce condition
$\dot{H}_0>0$ leads to the following relation
\begin{gather}
\frac{2\pi}{3M_p^2}\,\left(4V_0-\dot{\phi}_0^2\right)^{2}+
3\left(V_0'^2-V_0''\dot{\phi}_0^2\right)
+\frac{18V_0'^2\dot{\phi}_0^2}{\left(4V_0-\dot{\phi}_0^2\right)}>0
\tag{\ref{eq29}$'$}\label{eq29pr}
\end{gather}

The relation (\ref{eq29pr}) is compatible with the following
condition
\begin{equation}
4V_0-\dot{\phi}_0^2>0,
\end{equation}
which follows from discussed above restriction on the value of $Z$
in the case large (in module) negative value of $\beta$
($\beta<0$). Corresponding inflationary models are regular in
metrics, Hubble parameter and torsion (nonmetricity). (Note that
according to Eq. (24) the right-hand part of Eq. (26) tends to
$\frac{1}{2}H$ at $Z\to 0$). Unlike the condition (7$'$) of GR,
inequality (30) does not include radiation energy density
$\rho_r$. In contrast to GR, the appearance of VGRE in GTG (with
large negative $\beta$) does not depend on radiation, although the
contribution of ultrarelativistic matter to energy density of the
Universe at the bounce can be essentially greater than
contribution of scalar fields.

Let us consider cosmological equations of GTG (24)--(26) in
approximation of large negative $\beta$
($\left|\beta\left(4V-\dot{\phi}^2\right)\right|\gg 1$), by
supposing that
$\rho_r+\frac{1}{2}\dot{\phi}^2+V\ll\left|\beta\right|
\left(4V-\dot{\phi}^2\right)^2$.\footnote{This assumption does not
exclude that radiation energy density can dominate at the bounce
$\rho_r\gg V+\frac{1}{2}\dot{\phi}^2$.} We have
\begin{equation}
\frac{k}{R^2}+\frac{\left[2H\left(2V+
\dot{\phi}^2\right)+3V'\dot{\phi}\right]^2}{\left(4V-\dot{\phi}^2\right)^2}
=\frac{2\pi}{3M_p^2} \left(4V-\dot{\phi}^2\right),
\end{equation}
\begin{multline}
  \dot{H}\left(2V+\dot{\phi}^2\right)\left(4V-\dot{\phi}^2\right)+
  H^2\left(8V^2-34V\dot{\phi}^2-\dot{\phi}^4\right)-12H
  V'\dot{\phi}\left(2V+\dot{\phi}^2\right)\\
  + \frac{3}{2}\left(V''\dot{\phi}^2-V'^2\right)\left(4V-\dot{\phi}^2\right)-
  9V'^2\dot{\phi}^2
  =  \frac{\pi}{3M_p^2}\left(4V-\dot{\phi}^2\right)^3,
\end{multline}
\begin{equation}\
  S-\frac{1}{4}Q_1=-\frac{3}{2}\,
  \frac{V'+H\dot{\phi}}{4V-\dot{\phi}^2}\, \dot{\phi}\,.
\end{equation}
Eqs. (31)--(33) do not include radiation energy density, which is
not essential for dynamics of regular inflationary models in
considered case. According to Eq. (31) the scale factor at the
bounce  $R_0$ is determined from
\begin{equation}
\frac{k}{R_0^2}+9\left(\frac{V_0'\dot{\phi}_0}{4V_0-\dot{\phi}^2_0}\right)^2=
\frac{2\pi}{3M_p^2}\left(4V_0-\dot{\phi}^2_0\right).
\end{equation}
The value of $R_0$ determined by (34) coincides with that of GR
(formula (9)) only if $\rho_{r0}=0$ and $\dot{\phi}_0=0$.
According to Eq. (32) we have
\begin{align}
\dot{H}_0=&\left[\frac{\pi}{3M_p^2}\,\left(4V_0-\dot{\phi}_0^2\right)^{3}+
\frac{3}{2}\left(V_0'^2-V_0''\dot{\phi}_0^2\right)\left(4V_0-\dot{\phi}_0^2\right)
+9V_0'^2\dot{\phi}_0^2\right] \notag \\
&\times\left(2V_0+\dot{\phi}_0^2\right)^{-1}
\left(4V_0-\dot{\phi}_0^2\right)^{-1},
\end{align}
that corresponds to (29) in the case of large values of $|\beta|$.
In accordance with Eq. (34) for closed models the following
inequality at a bounce takes place
$$
\frac{2\pi}{3M_p^2}\left(4V_0-\dot{\phi}^2_0\right)^3-9 {V_0}'^2
\dot{\phi}^2>0,
$$
and in the case of open and flat models we have
$$
\frac{2\pi}{3M_p^2}\left(4V_0-\dot{\phi}^2_0\right)^3-9 {V_0}'^2
\dot{\phi}^2\leq 0.
$$
Because for open and flat models the derivative $\dot{\phi}$ does
not vanish, their evolution is essentially asymmetric with respect
to the bounce point.

At the beginning of the inflationary stage ($t\sim t_1$) the
Hubble parameter $H(t_1)$ from (32) is equal to
\begin{equation}
H(t_1)=\sqrt{\frac{8\pi}{3M_p^2}V_1+\frac{3}{4}\frac{V_1'^2}{V_1}\,}
\qquad \left(V_1=V(\phi(t_1)\right).
\end{equation}
In the case of linear time dependence of Hubble parameter during
stage of transition from compression to expansion we can simply
estimate the time interval $\Delta t_1$ by using relations
(35)--(36) and then determine the values of scale factor
$R_1=R_0(1+\dot{H}_0\Delta t_1)$ and scalar field
$\phi_1=\phi_0+\dot{\phi}_0\Delta t_1$ at the end of transition
stage (as it was made for inflationary models in GR in Section 2).
Numerical analysis  of regular inflationary models in GTG with
large $|\beta|$ shows that the most important difference of such
models in comparison with that of GR concerns the transition stage
from compression to expansion and the stage after inflation. As
was noted above the region of initial conditions at the bounce of
regular inflationary solutions is essentially more wide in GTG
than in GR (compare (7$'$) with (\ref{eq29pr}), (30)).
\begin{figure}[htb!]
\begin{minipage}{0.48\textwidth}\centering{
\epsfig{file=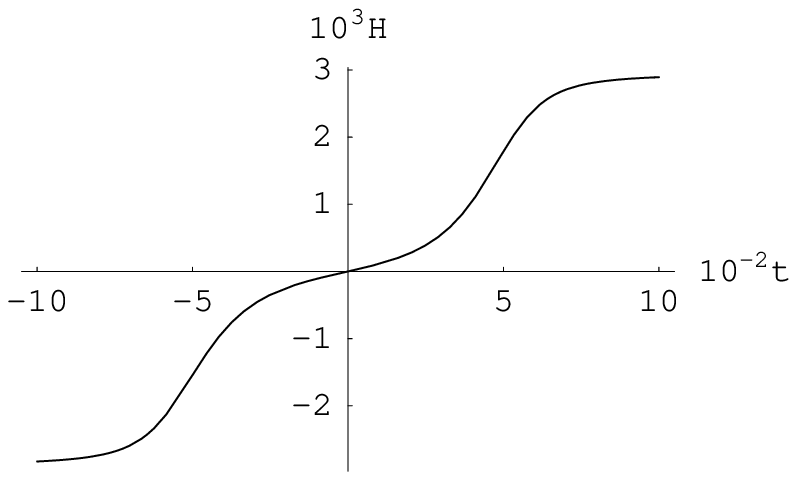,width=\linewidth}}
\end{minipage}\, \hfill\,
\begin{minipage}{0.48\textwidth}\centering{
\epsfig{file=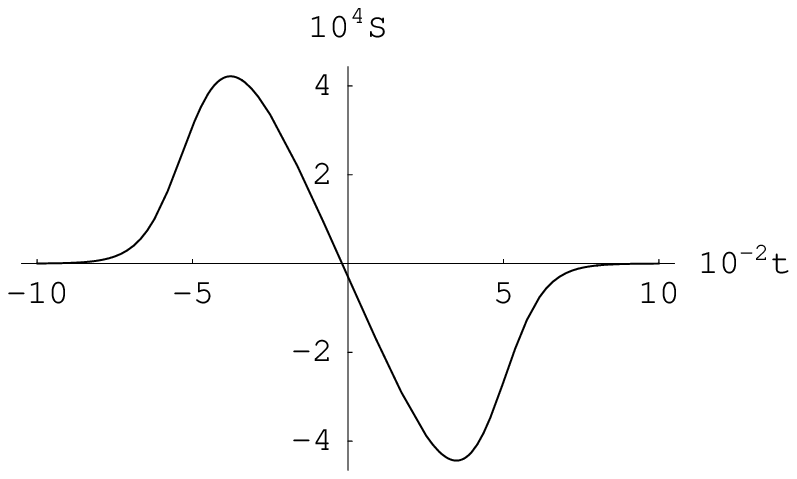,width=\linewidth}}
\end{minipage}\\
\begin{minipage}{0.48\textwidth}\centering{
\epsfig{file=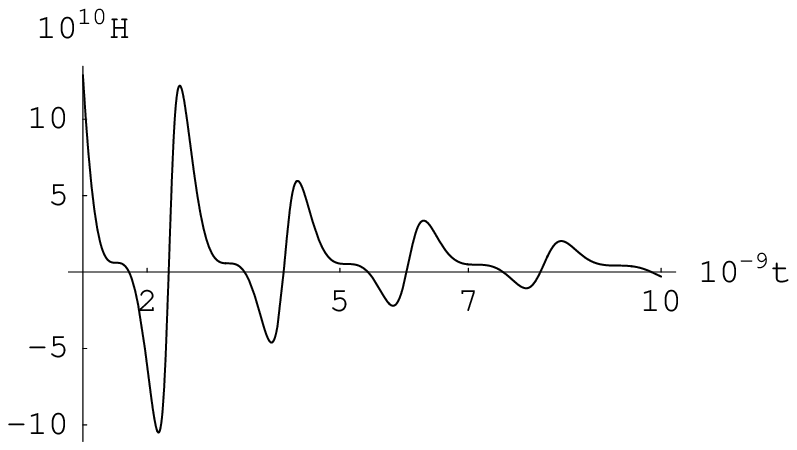,width=\linewidth}}
\end{minipage}\, \hfill\,
\begin{minipage}{0.48\textwidth}\centering{
\epsfig{file=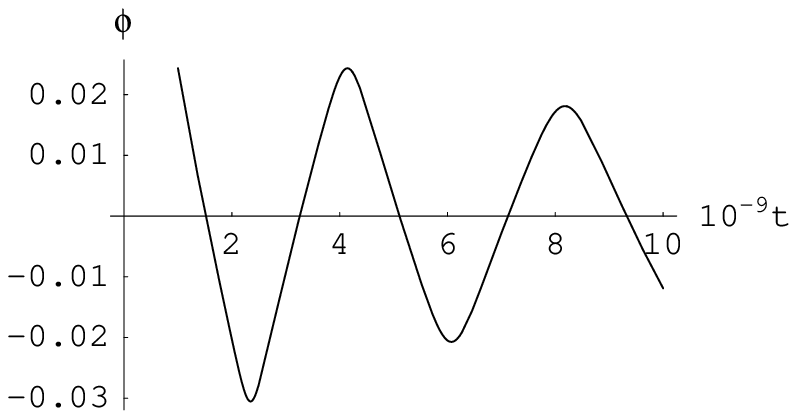,width=\linewidth}}
\end{minipage}\\
\begin{minipage}{0.48\textwidth}\centering{
\epsfig{file=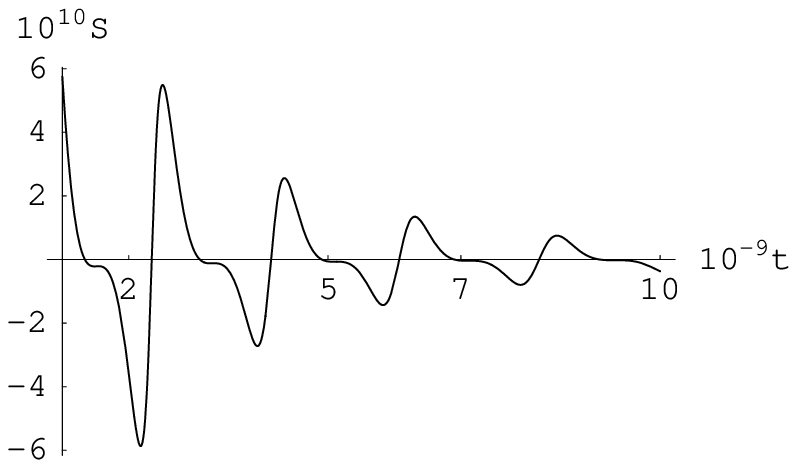,width=\linewidth}}
\end{minipage}\, \hfill\,
\caption[]{Dynamics of characteristics of inflationary model in
GTG with $\beta=-10^{20}M_p^{-4}$ during transition stage and
stage after inflation ($V_0=10^{-6}M_p^4$,
$\dot{\phi}_0=\sqrt{2V_0}$, $\rho_{r0}=10^5V_0$).}
\end{figure}
In Fig. 4 the evolution of different characteristics of regular
inflationary model in GTG ($\beta=-10^{20}M_p^{-4}$) is given by
choosing of initial conditions, by which a bouncing solution in GR
is impossible. From Fig. 4 we see that the Hubble parameter and
torsion as well as scalar field oscillate after inflation.
Amplitude and frequency of scalar field oscillations are smaller
in comparison with that of GR (for given potential) and decrease
by increasing of $|\beta|$.

\section*{Conclusion}

Considered regular inflationary models can be interesting for
inflationary cosmology, if the realization of physical conditions
leading to a bounce can take place at the end of cosmological
compression. In the frame of GR this means that the greatest part
of energy density of the Universe has to be determined by scalar
fields (see (7$'$)). However, in the framework of GTG the
realization of the bounce is possible, when usual
ultrarelativistic matter dominates at the end of cosmological
compression, and scalar fields satisfy more weak restriction
(relation (30)) --- its kinetic energy density can be greater than
the scalar field potential. Unlike GR, in the frame of GTG the
elimination of cosmological singularity and realization of the
bounce in inflationary models are ensured by cosmological
equations (24)-(25).

\section*{Acknowledgements}

I am grateful to my pupil Alexander Garkun and my son  Andrey
Minkevich for technical help by preparation of this paper. I am
thankful to my wife Eleonora for her support.

\end{document}